\documentclass[12pt,preprint]{aastex}

\shorttitle{Two types of dynamic structures} \shortauthors{Zhang
\& Li}

\begin{document}

\title{Two Types of Dynamic Cool Coronal Structures Observed
with STEREO and HINODE}

\author{Jun Zhang and Leping Li}

\affil{Key Laboratory of Solar Activity, National Astronomical
Observatories, Chinese Academy of Sciences, Beijing 100012, China;
zjun;lepingli@ourstar.bao.ac.cn}

\begin{abstract}

Solar coronal loops show significant plasma motions during their
formation and eruption stages. Dynamic cool coronal structures, on
the other hand, are often observed to propagate along coronal loops.
In this paper, we report on the discovery of two types of dynamic
cool coronal structures, and characterize their fundamental
properties. Using the EUV 304 {\AA} images from the Extreme
UltraViolet Imager (EUVI) telescope on the Solar TErrestrial
RElation Observatory (STEREO) and the Ca~II filtergrams from the
Solar Optical Telescope (SOT) instrument on HINODE, we study the
evolution of an EUV arch and the kinematics of cool coronal
structures. The EUV 304 {\AA} observations show that a missile-like
plasmoid moves along an arch-shaped trajectory, with an average
velocity of 31 km s$^{-1}$. About three hours later, a plasma arch
forms along the trajectory, subsequently the top part of the arch
fades away and disappears, meanwhile the plasma belonging to the two
legs of the arch flows downward to the arch feet. During the arch
formation and disappearance, SOT Ca II images explore dynamic cool
coronal structures beneath the arch. By tracking these structures,
we classify them into two types. Type I is threadlike in shape and
flows downward with a greater average velocity of 72 km s$^{-1}$,
finally it combines a loop fibril at chromospheric altitude. Type II
is shape-transformable and sometimes rolling as it flows downward
with a smaller velocity of 37 km s$^{-1}$, then disappears insularly
in the chromosphere. It is suggested that the two types of
structures are possibly controlled by different magnetic
configurations.

\end{abstract}


\section{INTRODUCTION}

The solar atmosphere is extremely complex and magnetically
structured. Coronal loops, magnetically closed structures in the
upper solar atmosphere, exhibit intrinsically dynamic behavior (De
Groof et al.~\cite{de04,de05}; Doyle et al.~\cite{doy06}; O'Shea et
al.~\cite{osh07a,osh07b}), and dynamic coronal structures are often
seen to propagate along these loops (Engvold et al.~\cite{eng79};
Loughhead \& Bray~\cite{lou84}; Brekke et al.~\cite{bre97}).
Material condensing in the corona and appearing to rain down into
the chromosphere as observed at the solar limb is termed as
``coronal rain", originally based on H$\alpha$ observations
(Tandberg-Hanssen~\cite{tan74}). Levine \& Withroe~(\cite{lev77})
presented observations of an active region loop system which
underwent a sudden disruption leading to the evacuation of most of
the material initially present below a temperature of about
2$\times$10$^6$ K. Kjeldseth-Moe \& Brekke (\cite{kje98}) studied
the temporal variability of active region loops with the Coronal
Diagnostic Spectrometer (CDS), and reported significant changes of
coronal loops over a period of one hour, in particular seen in
emission lines in the temperature range between 1-5$\times$10$^{5}$
K. This variability is accompanied by large Doppler shifts,
typically around 50-100 km s$^{-1}$. Observations of downflows above
large-scale post-flare arcades were first reported by McKenzie \&
Hudson (\cite{mck99}) using the soft X-ray observations. These
downflows are discribed as dark structures moving through the corona
towards the Sun with velocities in the range of 45-500 km s$^{-1}$.
Observations with the Extreme-ultraviolet Imaging Telescope (EIT)
with high temporal cadence furthermore reveal spatially localized
brightening in coronal loops, moving rapidly down towards the
footpoints of the loops (De Groof et al.~\cite{de04,de05}). Even
under quiescent condition, with no flares, loops show strong
temporal variability of emission in UV spectral lines and
substantial plasma flows. Schrijver~(\cite{sch01}) presented a
detailed study of ``catastrophic cooling" and evacuation of
quiescent coronal loops observed by Transition Region And Coronal
Explorer (TRACE), and shown that loop evacuation occurs frequently
after plasma in the upper parts of the loops has cooled to
transition region or lower temperatures. The cooling process is
often accompanied by emission in C\,{\sc iv} (1548 \AA) and Ly
$\alpha$, developing initially near the top of the loop.
Spectroscopic investigations of loops show that the intensity
variations have different signatures in UV spectral lines and
exhibit Doppler shifts of 20-130 km s$^{-1}$ (Fredvik et
al.~\cite{fre02}).

The physical explanation generally given for dynamic coronal
structures is that the plasma which was evaporated into the tops of
loops (Forbes \& Malherbe~\cite{for86}; Svestka et al.~\cite{sve98})
radiates away its energy and cools down. The fact that coronal loops
undergo rapid evacuation has been known for decades. Thereafter,
cool plasma is observed to slide down on both sides of the loop
toward the footpoints with velocities of up to 100 km s$^{-1}$.
Shimojo et al.~(\cite{shi02}) carried out a joint observations with
TRACE and Hida/Domeless Solar Telescope (DST) and concluded that the
coronal rains over the active region are produced by transient
coronal loop brightenings or microflares. Innes et
al.~(\cite{inn03a,inn03b}) reported similar features detected in EUV
by TRACE and SOHO/Solar Ultraviolet Measurements of Emitted
Radiation (SUMER). The downflow here was observed as a dark trail of
plasma moving towards the Sun at TRACE 195 \AA~and in the Fe XXI \&
UV continuum by SUMER. Based on these observations, Innes et al.
(\cite{inn03a,inn03b}) concluded that the dark trails of the plasma
are likely to be the plasma voids introduced by McKenzie \&
Hudson~(\cite{mck99}). Asai et al.~(\cite{asa04}) used the
observations of TRACE and RHESSI to identify a time correlation
between non-thermal radiation in the hard X-ray spectra and the
times of downflows, and concluded that the downflow motions occurred
when large amounts of magnetic energy was released, suggesting a
relation of downflows with reconnection outflows. Tripathi et
al.~(\cite{tri06}) studied a distinct downflow in the course of a
prominence eruption associated with CME, and provided support for
the pinching off of the field lines drawn-out by the erupting
prominences and the contraction of the arcade formed by the
reconnection.

In this paper, for the first time, we report two types of dynamic
cool coronal structures which display differently from three
aspects: falling manner, falling velocities and disappearing
patterns, based on high tempo-spatial observations. Other
activities, e.g. evolution of plasma clouds, are also described.

\section{OBSERVATIONS}

The Extreme UltraViolet Imager (EUVI) telescope on the Solar
TErrestrial RElation Observatory (STEREO, Howard et
al.~\cite{how08}) monitors the solar chromosphere and corona. The
detectors of EUVI have a field-of-view out to 1.7 $R_{\sun}$, and
observe in four spectral channels that span the 0.1 to 2.0 MK
temperature range. The Solar Optical Telescope (SOT) instrument on
Hinode (Tsuneta et al.~\cite{tsu08}), on the other hand, explores a
continuous, seeing-free series of diffraction-limited images in
3880-6680 {\AA} range with very high tempo-spatial resolutions. On
2007 May 10, there was an arch, which was located on the eastern
limb (S02E90) of the Sun, and observed by the EUVI. In this study we
employ the EUV 304 {\AA} (60,000-80,000 K) images to investigate the
evolution of the arch. The observational mode of the 304 {\AA}
images is 1.59$''$ per CCD pixel with a cadence of 38 s. The Ca II
(10,000-20,000 K) filtergrams from SOT detected a sub-area
underneath the arch. These filtergrams have a spatial resolution of
0.22$''$, and with a cadence of 20 s. Figure 1 displays an EUV 304
{\AA} image from EUVI ({\it left}) and a Ca II image from SOT ({\it
right}).

\subsection{Dynamics of The Arch}

Solar coronal arches (or loops) can be detected when observed in
coronal UV and EUV lines (Zhang \& Wang~\cite{zha00,zha01}; Anzer et
al,~\cite{anz07}). During the formation stage of coronal arches,
significant plasma motions are always observed. Similarly, mass
motions are also distinct during the disappearance of loops. Figure
2 shows time sequence of the EUV 304 {\AA} images from EUVI. The top
row displays the motion of a missile-like plasmoid (denoted by the
arrows). The plasmoid moved along an arch-shaped trajectory. From
17:49 UT to 18:08 UT, when the plasmoid rose to the apex of the arch
trajectory, the moving velocities decreased from 35 to 24 km
s$^{-1}$. Then the plasmoid fell downward from the apex. The falling
velocities increased monotonously, at 18:46 UT, the velocity reached
36 km s$^{-1}$. To reduce the error in the determination of these
velocities, we measure each velocity in a time interval of about 10
minutes. The uncertainty of the plasmoid position (one pixel,
1.59$''$) results in a velocity error of 2 km s$^{-1}$. Almost three
hours later, plasma brightened continuously along the former
plasmoid trajectory, the arrows in the bottom row denote the front
edge of the brightening plasma. At 22:53 UT the plasma inside the
other leg of the arch also brightened to meet the foregoing plasma,
thus making the plasma arch connecting the two leg, with a length of
${\sim}$200 Mm. Subsequently the material in the top part of the
arch fades away, the material within the two legs moved downward to
the two feet of the arch, then the arch disappeared in 20 minutes.

\subsection{Kinematics of Cool Coronal Structures and Plasma Clouds underneath The
Arch}

From 2007 May 10, 21:20 to 23:59 UT, the target of the SOT Ca II
images was the eastern solar limb region located underneath the arch
(see Fig. 1). Dynamic structures and plasma clouds were detected
during the observations. Figure 3 presents the evolution of two
structures from the Ca II filtergrams. The left column displays the
falling process of the first structure (see the arrows) which was
initially detected at 22:05:00 UT. This structure appeared as a
threadlike (i.e., elongated) in shape, and fell directly downward to
the solar chromosphere. In order to compare with the free-fall
tracks, we used the height of the falling structures to the solar
surface to calculate the falling velocity. Furthermore, all the
structures were generally falling vertical to the solar surface. The
centers of the gravity of the structures are considered as their
position. Under this condition, we get a velocity of 75 km s$^{-1}$.
At 22:07:19 UT, the structure split into two segments (see the two
arrows). The two segments continuously fell downward with a nearly
constant velocity. Just before they met two detectable loop fibil
(indicated by two dotted lines at 22:08:42 UT), their velocities
decreased from 75 to 60 km s$^{-1}$.

The evolution of the second dynamic structure is shown in the right
column of Fig. 3. The difference between these two structures mainly
manifests on three aspects. The first one is the falling manner. The
shape of this structure changed continuously as it falls downward.
Sometimes the shape is threadlike (indicated by dotted lines), and
sometimes a drop (see the arrows). It is noticed that the structure
rotated about 50$^{\circ}$ from 22:22:41 to 22:25:41 UT, before the
structure split into two parts. The second one is falling velocity.
From 22:22:41 to 22:26:21 UT, the velocity of the dynamic structure
was 52 km s$^{-1}$ on average, about 20 km s$^{-1}$ lower than the
falling velocity of the first structure, then it decreased to 38 km
s$^{-1}$. The last aspect is the disappearing patterns. This
structure disappeared insularly at a height of 5.5 Mm over the solar
surface, no loop was detected beneath the disappearing position of
the structure.

Based on their disappearing patterns, we classify these dynamic
structures into two types. Type I includes 16 isolated structures
which disappear by merging with loop fibril. These structures are
tracked while they falling downward from corona to chromosphere.
Type II includes 10 isolated structures which disappear insularly in
the chromosphere. They are also tracked during their falling
downward. Figure 4 displays the relationship between the falling
velocities of dynamic structures and the height to the solar
surface. The solid circles represent the 16 structures of type I,
and the hollow circles, the 10 structures of type II. It indicates
that the type I structures fall at greater velocities, with an
average value of 72 km s$^{-1}$. However, the type II structures
move slowly downward with an average velocity of 37 km s$^{-1}$,
which is only a half falling velocity of the type I structures.
Comparing the observed falling velocities with calculated free-fall
tracks, we find that all the dynamic structures deviate distinctly
from free-fall tracks, while they are close to the solar limb.

The measured falling velocities of these dynamic structures only
represent the projective component of true velocities, they are
affected by the angle between the moving direction of these
structure and the line-of-sight of the observer. De Groof et
al.~(\cite{de04}) have discussed the projection effect.
Theoretically the projection effect (see Fig. 10 of De Groof et
al.~\cite{de04}) can not cause obvious speed deviation, while all
the dynamic structures are close to the limb. The main error source
of velocity determination is due to uncertainty in the dynamic
structure position. A position error of one pixel introduces an
error of the velocity of (one pixel)/(time interval). As the size of
one pixel is about 0.11\arcsec~and the time interval is about 20
seconds, the error in velocity becomes 4 km s$^{-1}$, which may
explain a part of the scatter in Fig. 4. However it is clear that
the velocity error is much smaller than the falling velocity
difference (35 km s$^{-1}$) of the two types of dynamic structures.
This implies that the velocity difference exists in truth.

Besides the individual dynamic structures, there are other kinds of
plasma structures. Figure 5 shows the evolution of cloud-built
plasma. The plasma within each ellipse region fall as a whole
downward to the chromosphere, but the velocities are not unique.
There are some bright structures (see the arrows), they become
larger and larger in size during their falling downward. The average
velocity of these structures is about 11 km s$^{-1}$. In the
observational interval, there are two such cases. Considering the
size of these structures, we speculate that they may correspond with
coronal rains reported before (Zirin~\cite{zir74}). Some plasma
clouds, however, do not fall downward. Figure 6 shows a cyclone-like
motion of a plasma cloud. The dotted curves represent the central
axes of the cloud at different evolution stages. At 22:37:22 UT the
axis was nearly vertical to the solar surface, then the axis bent at
the top part to form a fishhook-shaped structure. This structure
became larger and larger in size, then diffused and disappeared in
half an hour.

\section{Conclusions and Discussion}

We first report on the observations of two types of dynamic
structures underneath an arch. Type I structures fall downward with
greater velocities, and disappear by combining detectable loop
fibril. However type II ones fall with smaller velocities, then
disappear insularly in the chromosphere. It has been identified that
all the dynamic structures deviated remarkably from free-fall
tracks, while they are close to the solar limb. In additional, other
kinds of plasma structures, e.g. cloud-built plasma, also exist in
the corona and upper chromosphere. These plasma clouds either move
downward to the lower chromosphere, or display cyclone-like motion.

The difference of the falling velocities between these two types of
structures may be affected by the magnetic configuration, as showed
in Fig. 7. From this figure, we can see that type I structures fall
along slick magnetic loops which are mainly perpendicular to the
solar surface. For type II structures, however a more complicated
magnetic configuration may respond for their strange kinematics,
i.e. undergoing rotation and falling along tortuous paths instead of
straight down. For better comparison with the observations displayed
in Fig. 3, we denoted the actually observational times (shown in
Fig. 3) in Fig. 7. Malville~(\cite{mal76}) has also suggested that
material falling in distorted flux ropes experiences a Lorentz
force, thus makes the structures fall slower.

The departures of the dynamic structures from free-fall at lower
heights was interpreted in terms of plasma deceleration (Wiik et al.
\cite{wii96}). Heinzel et al.~(\cite{hei92}) have suggested three
possible explanations for the observed deceleration: variation of
the loop magnetic structure during the typical fall-time of the
dynamic structures; helicity of the magnetic field inside the loop;
and deceleration of cool structures as they move inside a hotter
plasma. In this work, whether the first explanation is appropriate
for the departure is still in question, as the variation of loops
can not be reliably detected in our observations. The last
explanation can be considered at lower heights where cool structures
compress a hot plasma as in a piston. Using MHD simulations, Mackay
\& Galsgaard~(\cite{mac01}) have considered falling high density
material in a stratified atmosphere, they produced a piston type
model to interpret the deceleration of plasma. For the fast falling
dynamic structures, the piston mechanism may play a key role, as the
bright (hot and/or denser) plasma at the footpoints of the fibrils
is detected (see Fig. 3). For the slow falling ones, the second
explanation is possible, as there exists some evidence that these
structures move along twisted magnetic loops. Mein et
al.~(\cite{mei96}) observed similar departures from free-fall in the
case of downflows in arch filament systems, and proposed another
good candidate for deceleration. They suggested that a deceleration
may result from shock waves generated at the footpoints, presumably
due to the interaction of downflows with denser parts of the
atmosphere.

The formation of these dynamic structures is also an intriguing
problem. Generally, dynamic cool coronal structures are the result
of the cooling of the hot coronal loops which we typically observe
at somewhat higher altitudes. Shimojo et al.~(\cite{shi02}) once
concluded that the coronal rains over the active region are produced
by transient coronal loop brightenings or microflares. M\"{u}ller et
al.~(\cite{mul03,mul05}) presented a comparison of observed
intensity enhancements from an EIT shutterless campaign with
non-equilibrium ionization simulations of coronal loops in order to
reveal the physical processes governing fast flows and localized
brightenings. They show that catastrophic cooling around the loop
apex as a consequence of footpoint-concentrated heating can perhaps
offer a simple explanation for these observations. In this paper,
the cool dynamic structures also appeared under an dynamic EUV arch,
they may relate with the formation and disappearance of the arch.
Further study of dynamic structures will require new optical and UV
spectral observations with high resolution, this would allow us to
make possibly a reliable geometrical reconstruction of true loop
shapes, and to derive the plasma parameters more accurately.

\acknowledgements

The authors are indebted to the {\it STEREO} and {\it Hinode} teams
for providing the data. {\it Hinode} is a Japanese mission developed
and launched by ISAS/JAXA, with NAOJ as domestic partner and NASA
and STFC (UK) as international partners. It is operated by these
agencies in co-operation with ESA and NSC (Norway). This work is
supported by the National Natural Science Foundations of China
(G40890161 and 40674081), the CAS Project KJCX2-YW-T04, and the
National Basic Research Program of China under grant G2006CB806303.

\clearpage

\begin{figure}
\centering
\includegraphics[width=\textwidth, angle=0]{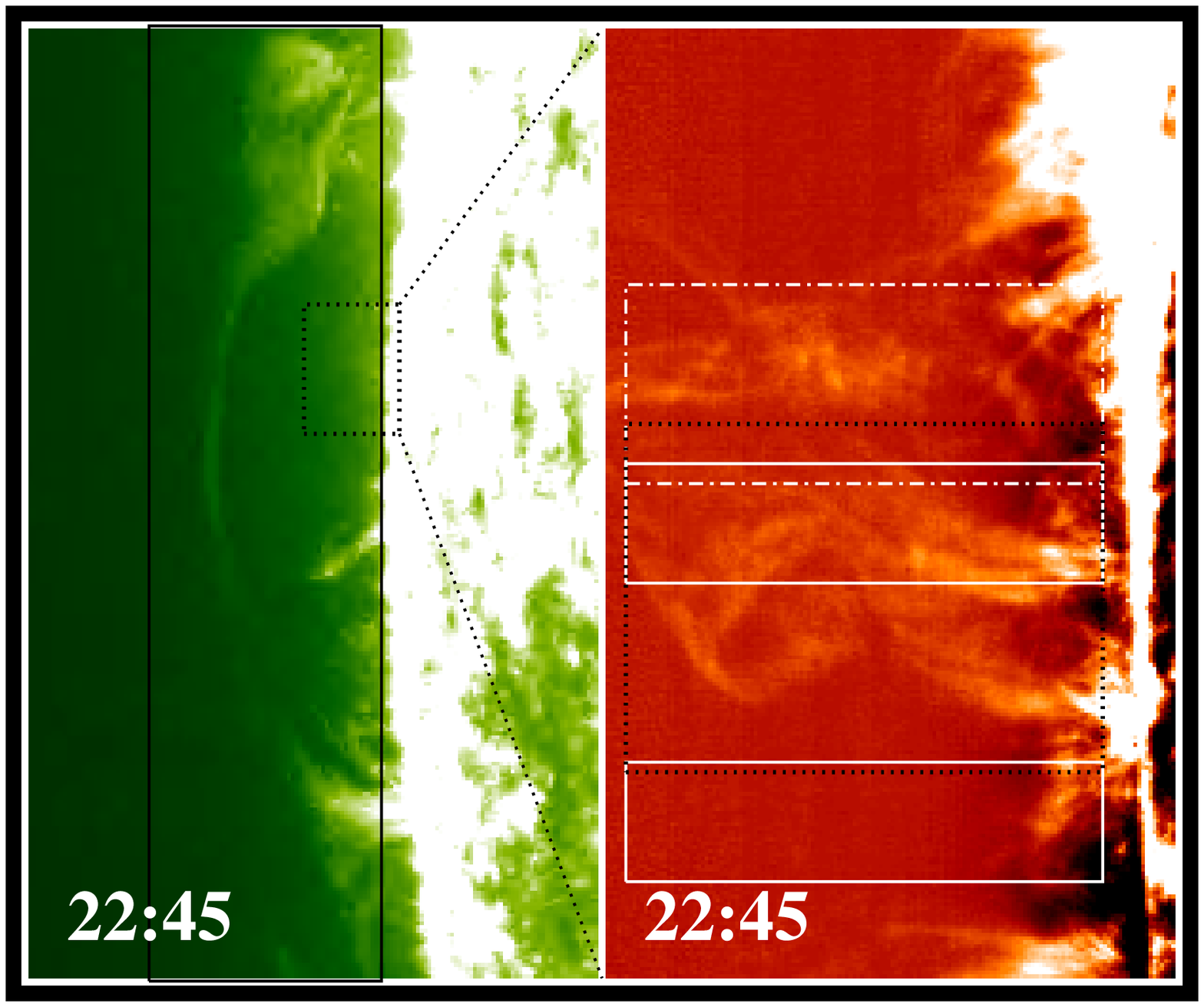}
\caption{An EUV 304 {\AA} image ({\it left}) from EUVI on STEREO and
a Ca II image ({\it right}) from SOT aboard Hinode on 2007 May 10.
On the EUV 304 {\AA} image, the larger box encloses an arch (see
Fig. 2), and the smaller box underneath the arch, outlines the
field-of-view (FOV) of the right Ca II filtergram.  The two solid
rectangle windows on the Ca II image denote sub-areas where
individual dynamic structures flow downward from corona to
chromosphere (see details in Fig. 3), the dash-dotted window, the
area where plasma shows waterfall-like motion (see also Fig. 5), and
the dotted window, the region in which plasma displays cyclone-like
motion (see Fig. 6). The FOV for the EUVI image is about
238$\arcsec$${\times}$397$\arcsec$, and
33$\arcsec$${\times}$55$\arcsec$ for the SOT Ca II filtergram. }
\label{Fig:fig1}
\end{figure}

\begin{figure}
\centering
\includegraphics[width=\textwidth, angle=0]{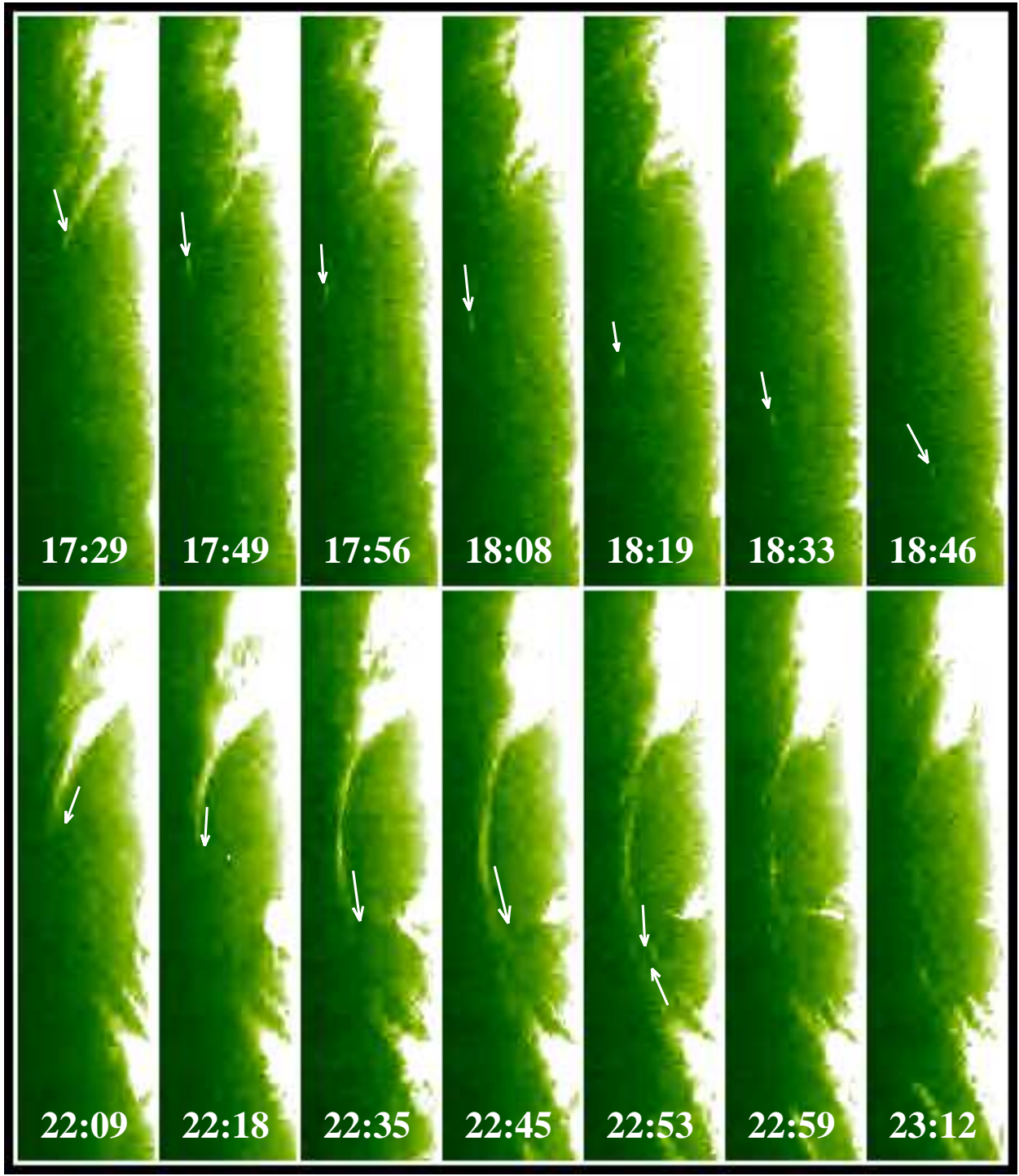}
\caption{Time sequence of the EUV 304 {\AA} images (see the larger
rectangle window in the left panel of Fig. 1) showing the dynamics
of the arch. The arrows are described in the text. The FOV is about
95$\arcsec$${\times}$397$\arcsec$.} \label{Fig:fig2}
\end{figure}

\begin{figure}
\centering
\includegraphics[width=\textwidth, angle=0]{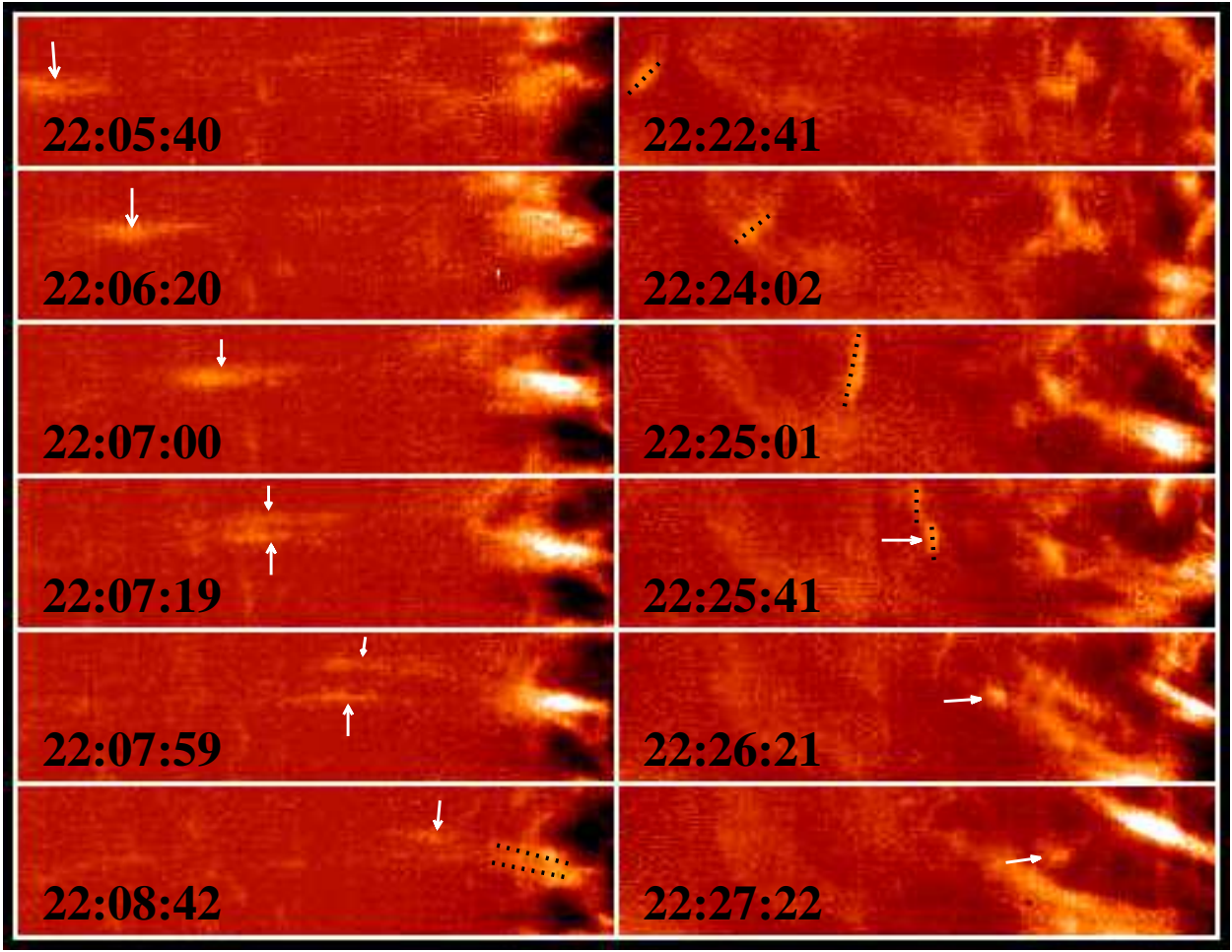}
\caption{Time sequence of the Ca II filtergrams showing the
evolution of two dynamic structures. The left column corresponds to
the region of the lower solid rectangle window in the right panel of
Fig. 1, and the right column to the region of the upper solid
rectangle window. The FOV is about
26.4$\arcsec$${\times}$6.6$\arcsec$. The arrows and dotted lines are
described in the text.} \label{Fig:fig3}
\end{figure}

\begin{figure}
\centering
\includegraphics[width=\textwidth, angle=0]{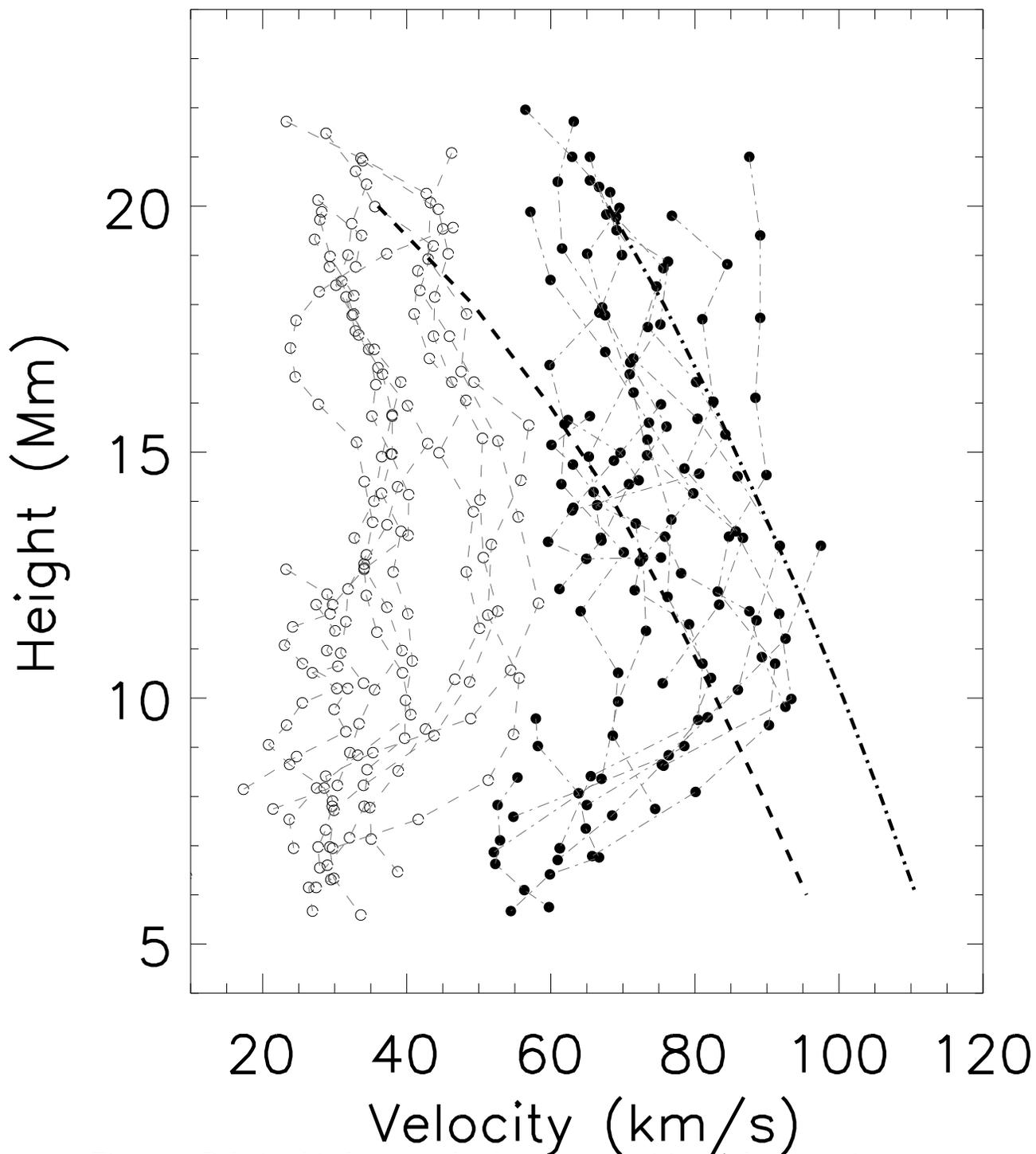}
\caption{Relationship between the downflow velocities of the dynamic
structures and the height to the solar surface. The solid circles
and the hollow circles denote the velocities of the two types of
dynamic structures (see the text), respectively. The individual
trajectories (the dash-dotted curves connecting the solid circles,
and dashed curves the hollow circles) of the 26 dynamic structures
are plotted. The two heavy curves representing the free-fall tracks
are fitted to the observed falling velocities at the heights where
the bright dynamic structures are detected.} \label{Fig:fig4}
\end{figure}

\begin{figure}
\centering
\includegraphics[width=\textwidth, angle=0]{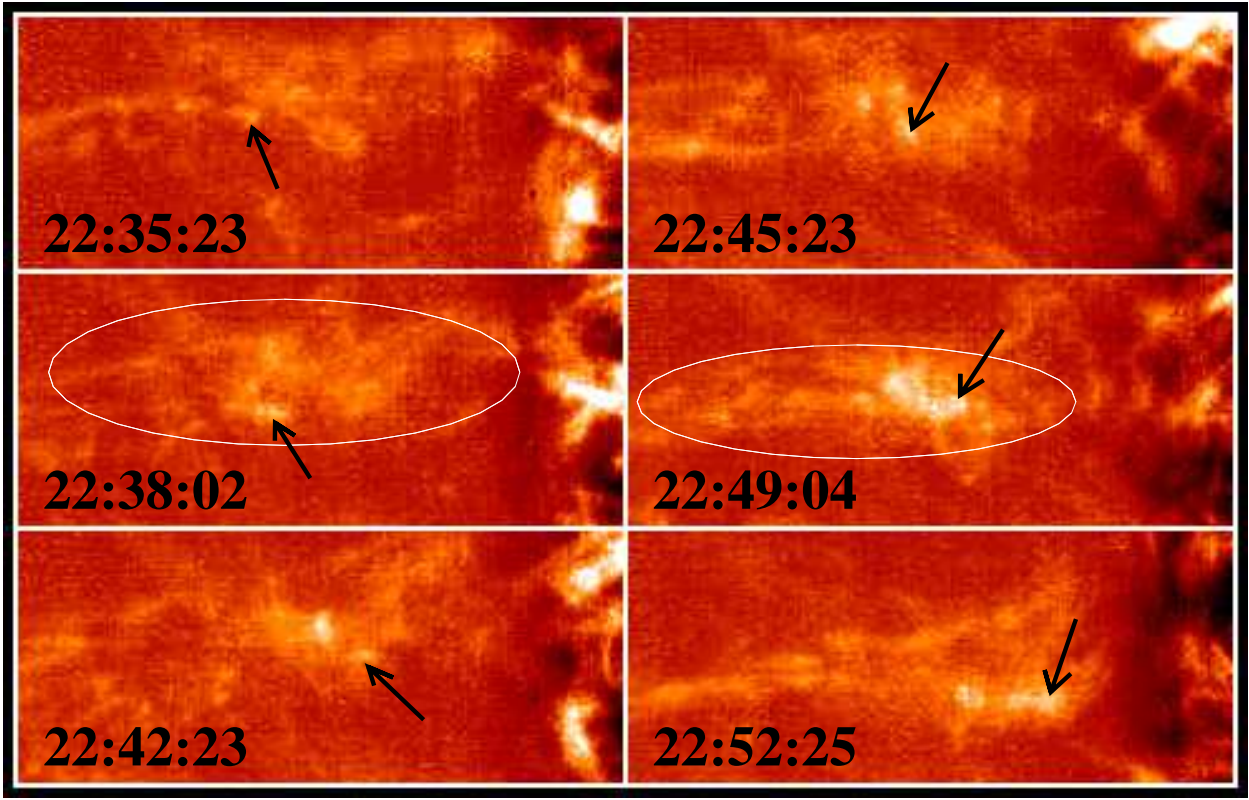}
\caption{Time sequence of the Ca II filtergrams (see the dash-dotted
window in the right panel of Fig. 1) showing the evolution of
waterfall-like plasma. The FOV is about
26.4$\arcsec$${\times}$11$\arcsec$. The arrows and two ellipses are
described in the text.} \label{Fig:fig5}
\end{figure}

\begin{figure}
\centering
\includegraphics[width=\textwidth, angle=0]{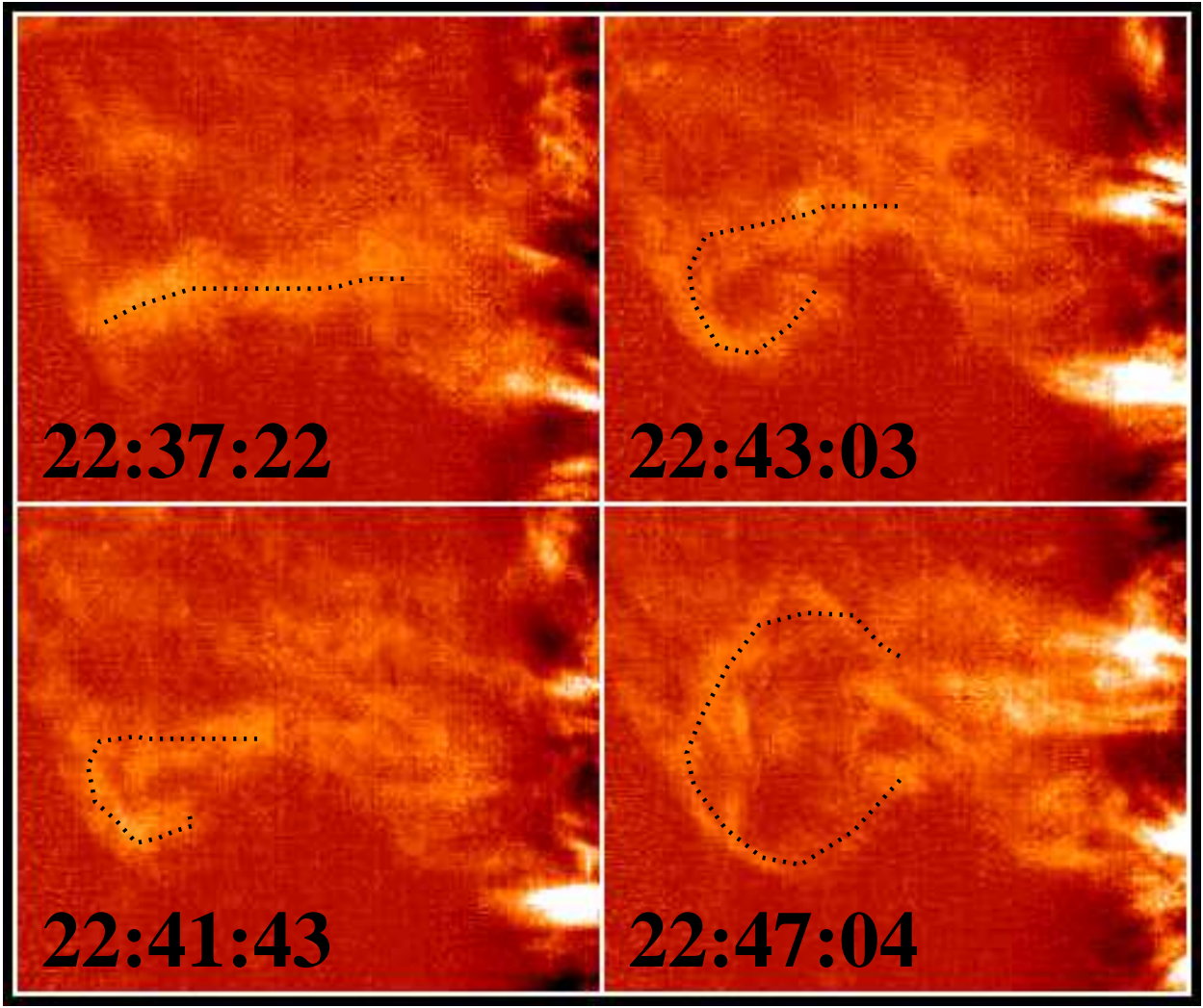}
\caption{Time sequence of the Ca II filtergrams (see the black
dotted window in the right panel of Fig. 1) showing the cyclone-like
motion of plasma. The FOV is about
26.4$\arcsec$${\times}$22$\arcsec$. The curves are described in the
text.} \label{Fig:fig6}
\end{figure}

\begin{figure}
\centering
\includegraphics[width=0.8\textwidth, angle=0]{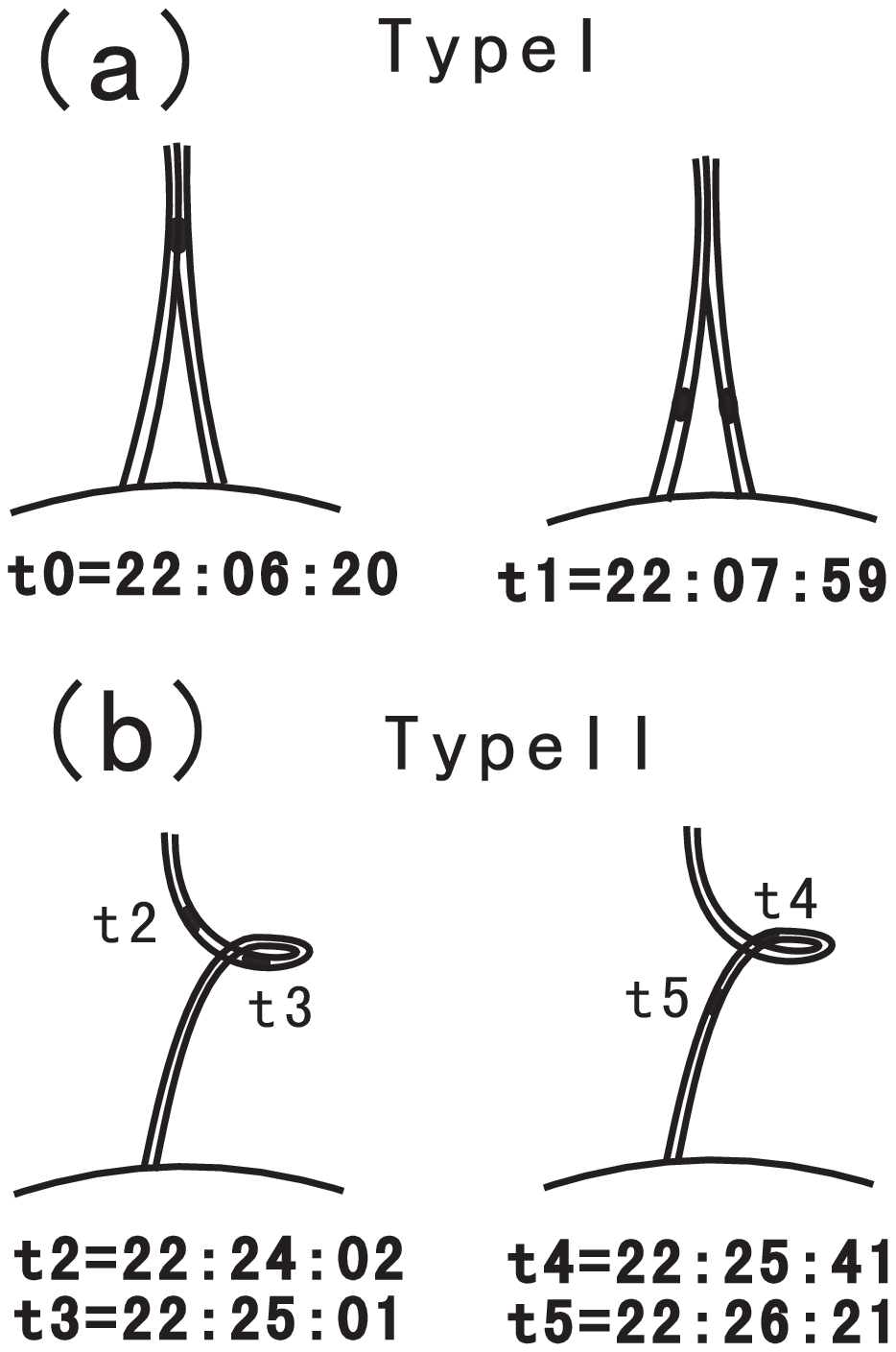}
\caption{The schematic diagrams illustrating the evolution of the
type I dynamic coronal structures at t0 and t1 (a), and type II
dynamic coronal structures at t2, t3, t4, and t5 (b), respectively.
t0, t1, t2, t3, t4, and t5 are the observational times displayed in
Fig. 3. The solid lines represent the magnetic field lines, as well
as the solid points, the dynamic coronal structures.}
\label{Fig:fig7}
\end{figure}

\end{document}